\documentstyle[12pt,epsf]{article}
\renewcommand{\bar}[1]{\overline{#1}}

\begin{document}

\begin{flushright}
USM-TH-137
\\ CPT-2003/P.4529
\end{flushright}
\bigskip\bigskip

\centerline{
\bf Transverse single-spin asymmetries in gauge boson production}

\vspace{22pt} \centerline{\bf Ivan Schmidt\footnote{e-mail:
ischmidt@fis.utfsm.cl}$^{a}$, Jacques Soffer\footnote{e-mail:
Jacques.Soffer@cpt.univ-mrs.fr}$^{b}$}

\vspace{2pt}

{\centerline {$^{a}$Departamento de F\'\i sica, Universidad
T\'ecnica Federico Santa Mar\'\i a,}}

{\centerline {Casilla 110-V, 
Valpara\'\i so, Chile}

{\centerline {$^{b}$Centre de Physique Th$\acute{\rm{e}}$orique,
CNRS, Luminy Case 907,}}

{\centerline { F-13288 Marseille Cedex 9, France}}

\vspace{4pt}
\begin{center} {\large \bf Abstract}

\end{center}

Transverse single-spin asymmetries (SSA) in inclusive reactions are
now considered to be directly related to the transverse momentum $k_{T}$
of the fundamental partons involved in the process. Several possible leading-twist
QCD mechanisms have been proposed to explain the
available data, in particular
the Sivers effect which was resurrected recently. We show that from
the measurement of transverse SSA in inclusive production of gauge bosons, one
can learn more about the Sivers functions and possibly to achieve a
reliable flavor separation.

\centerline{PACS numbers: 13.87.Fh, 13.60.Hb, 13.65.+i, 13.85.Ni}

\newpage

The existence of puzzling and significant experimental results for
transverse single-spin asymmetries (SSA) in several high energy
processes, has been known for a long time. Let us mention in
particular, the SSA in hyperon (and antihyperon) inclusive
production $pN\rightarrow Y^{\uparrow}X$, at various energies
\cite{LP}, and in pion inclusive production $p p^{\uparrow}
\rightarrow \pi X$ with a center-of-mass (c.m.) energy $\sqrt{s}
\sim 20 \mbox {GeV}$ \cite{E704}, an effect which seems to survive
at $\sqrt{s} = 200 \mbox {GeV}$, as observed for $\pi^{0}$
production, in the first spin run at BNL-RHIC \cite{GR}. For the
large amount of significant effects accumulated in the hyperon
production, in particular $\Lambda$'s, many theoretical attempts
to explain them have been proposed \cite{Felix, JS99}, with strong
limitations in most cases, so it is clear that this polarization
phenomena is not easy to understand within the pQCD framework.
Recently an azimuthal asymmetry has been also observed in
semi-inclusive deep-inelastic scattering (SIDIS) $ {\it
l}p^{\uparrow} \rightarrow {\it l}\pi X$, for targets polarized
transversely ($A_{UT}$) and longitudinally ($A_{UL}$) to the
direction of the unpolarized incoming lepton beam direction
\cite{SMC,hermes}. In a kinematic region where the generalized QCD
parton model is valid, the description of such a reaction involves
both distribution functions $q(x)$ and fragmentation functions
$D_q^{\pi}(z)$, where the scaling variable has been omitted
purposely. However in a phenomenological approach based on the
generalization of the factorization theorem, one can replace
$q(x)$ by $q(x,{\bf k}_T)$, with an explicit dependence on ${\bf
k}_T$, the transverse momentum of the quark inside the nucleon and
similarly for the fragmentation function, replacing $D_q^{\pi}(z)$
by $D_q^{\pi}(z,{\bf k}_T^{\pi})$, where ${\bf k}_T^{\pi}$ is the
transverse momentum of the $\pi$ with respect to the direction of
the fragmenting quark. When dealing with polarized processes, the
introduction of this transverse momentum dependence opens the way
to possible new spin effects, namely the Sivers effect \cite{DS},
related to the distribution function and the Collins effect
\cite{JCC}, related to the fragmentation function. More precisely,
the Collins function describes an azimuthal asymmetry in the
hadronization process of a transversely polarized quark, while the
Sivers function does the same for the production of a quark form a
transversely polarized hadron, usually the proton. The Sivers
effect was first discarded in Ref. \cite{JCC} on the basis of time
reversal invariance. However it was shown recently in Ref.
\cite{BHS1} that the observed asymmetry in SIDIS mentioned above,
which was until then attributed only to the Collins effect, could
be also explained by an effective Sivers effect. This mechanism is
due to QCD final-state interactions from gluon exchange between
the struck quark and the proton spectators, and it survives in the
Bjorken limit. Collins himself has reconsidered his proof of the
vanishing of the Sivers function \cite{JCC1} and it was shown in
Ref. \cite{JY} that the model results of Ref. \cite{BHS1} are
recovered properly in the light-cone gauge. In Drell-Yan
processes, QCD initial-state interactions from gluon exchange
between the incoming annihilating antiquark and the target
spectators, can also give rise to a non-zero Sivers function
\cite{BHS2} \footnote{ For a phenomenological study of transverse
SSA in Drell-Yan processes, see also Ref. \cite{BM,ADM}.}. The SSA
predicted for the Drell-Yan process $ \pi p^{\uparrow}\rightarrow
{\it l^{+}}{\it l^{-}} X$ is similar and of opposite sign to the
SSA in $ {\it l}p^{\uparrow}\rightarrow {\it l}\pi X$
\cite{JCC1,BHS2}.

Once we have identified these two fundamental leading-twist QCD
mechanisms to generate SSA, the Sivers and Collins effects, in
order to study them in more detail it is important to be able to
discriminate between them. \footnote{ The interference
fragmentation function with the measurement of two pions in the
final state is another possible leading-twist mechanism to produce
non-zero SSA in several processes \cite{CHL,JJT}.}. One way to achieve
this is to consider weak interaction processes, as proposed in
Ref. \cite{BHS3}. We recall that with the Collins mechanism, the
SSA is obtained from the transversity distribution function
$h_1^q$ of a quark of the initial polarized hadron, convoluted
with the Collins, ${\bf k}_T$ - dependent fragmentation function.
However in weak interaction processes, such as neutrino DIS on a
polarized target or $W$ production in polarized hadron-hadron
collisions, since the charged current only couples to quarks of
one chirality, $h_1^q$ decouples and thus the observed SSA is not
due to the Collins effect. This is not the case for the Sivers
effect which will be able to generate a non-zero SSA, as we will
see now.

Let us consider the inclusive production of a $W^{+}$ gauge boson in the
reaction $pp^{\uparrow}\rightarrow W^{+}X$, where one proton beam
is transversely polarized. In the Drell-Yan picture in terms of the
dominant quark-antiquark fusion reaction, the unpolarized cross-section reads
\begin{equation}
d\sigma=\int dx_ad^2{\bf k}_{Ta}dx_bd^2{\bf k}_{Tb}
[u(x_a,{\bf k}_{Ta})\bar d(x_b,{\bf k}_{Tb)}+ (u \leftrightarrow \bar d)]
d\hat \sigma^{ab\rightarrow W^{+}}.
\end{equation}
Similarly, the SSA defined as
\begin{equation}
A_N^{W^+} =\frac{ d\sigma^{\uparrow} - d\sigma^{\downarrow}}
{d\sigma^{\uparrow} + d\sigma^{\downarrow}}
\end{equation}
can be expressed such as
\begin{eqnarray}
d\sigma A_N^{W^+} &=& \int dx_ad^2{\bf k}_{Ta}dx_bd^2{\bf k}_{Tb}
[\Delta^{N}u(x_a,{\bf k}_{Ta})\bar d(x_b,{\bf k}_{Tb})
\nonumber \\
&& -\Delta^{N}\bar d(x_a,{\bf k}_{Ta})u(x_b,{\bf k}_{Tb}) ]
d\hat \sigma^{ab\rightarrow W^{+}}.
\end{eqnarray}
We recall the general definition of the ${\bf k}_T$ -dependent parton
distributions inside a transversely polarized proton, {\it up} labeled
with $\uparrow$ or {\it down} with $\downarrow$, namely
\begin{eqnarray}
q(x,{\bf k}_T)&=&
\frac{1}{2}[q_{\uparrow}(x,{\bf k}_T) + q_{\downarrow}(x,{\bf k}_T)]
\nonumber \\
&& = \frac{1}{2}[q_{\uparrow}(x,{\bf k}_T) + q_{\uparrow}(x,-{\bf k}_T)]=
 q(x,k_T) ,
\end{eqnarray}
whereas for the Sivers functions \cite {DS} we have
\begin{eqnarray}
\Delta q^N(x,{\bf k}_T)&=&
q_{\uparrow}(x,{\bf k}_T) -  q_{\downarrow}(x,{\bf k}_T)
\nonumber \\
&& = q_{\uparrow}(x,{\bf k}_T) -  q_{\uparrow}(x,-{\bf k}_T)=
\Delta q^N(x,k_T){\bf S}_p\cdot{\bf \hat{p}}\times {\bf k}_T.
\end{eqnarray}
Here ${\bf S}_p$ denotes the transverse polarization of the
proton of three-momentum ${\bf p}$ and ${\bf \hat {p}}$ is a unit vector 
in the direction of ${\bf p}$. A priori the ${\bf k}_T$
-dependence of all these parton distributions is unknown, but as a
first approximation one can assume a simple factorized form for
the distribution functions and take for example, as in
Ref.\cite{ADM},
\begin{equation}
q(x,k_T)= q(x)f(k_T),
\end{equation}
where $f(k_T)$ is flavor independent, and a similar expression for
the corresponding Sivers functions. In such a situation, it is
clear that the SSA will also factorize and then it reads
\begin{equation}
A_N^{W^+} (\sqrt{s},y,{\bf p}_T)= H(p_T)A^{+}(\sqrt{s},y){\bf S}_p\cdot{\bf \hat{p}}\times {\bf {p}}_T,
\end{equation}
where ${\bf p}_T$ is the transverse momentum of the $W^+$
produced at the c.m. energy $\sqrt{s}$ and $H(p_T)$ is a function
of $p_T$, the magnitude of ${\bf p}_T$. Obviously if the outgoing
$W^+$ has no transverse momentum, the SSA will be zero , as
expected. In the above expression we have now
\begin{equation}
A^{+}(\sqrt{s},y)=\frac{\Delta^{N}u(x_a)\bar d(x_b) - \Delta^{N}\bar d(x_a)u(x_b)}
{u(x_a)\bar d(x_b) + \bar d(x_a)u(x_b)},
\end{equation}
where $y$ denotes the $W^+$ rapidity, which is related to $x_a$
and $x_b$. Actually we have $x_a =\sqrt{\tau} e^y$ and  $x_b
=\sqrt{\tau} e^{-y}$, with $\tau =M_W^{2}/s$, and we note that a
similar expression for $A_N^{W^-}$, the SSA corresponding to $W^-$
production, is obtained by permuting $u$ and $d$. For the
$y$-dependent part of the SSA, one gets for $y=0$
\begin{equation}
A^{+}=\frac{1}{2}(\frac{\Delta^{N}u}{u} - \frac{\Delta^{N}\bar d}{\bar d})  \  \  \  \ \mbox{and} \  \  \ \
A^{-}=\frac{1}{2}(\frac{\Delta^{N}d}{d} - \frac{\Delta^{N}\bar u}{\bar u})
\end{equation}
evaluated at $x = M_W /\sqrt{s}$.
Moreover a real flavor separation can be obtained away from $y=0$, since for $y=-1$ one has
\begin{equation}
A^{+}\sim - \frac{\Delta^{N}\bar d}{\bar d}  \  \   \  \ \mbox{and}  \  \  \  \
A^{-}\sim - \frac{\Delta^{N}\bar u}{\bar u}
\end{equation}
evaluated at $x=0.059$ and for $y=+1$ one has
\begin{equation}
A^{+}\sim  \frac{\Delta^{N}u}{u} \  \   \  \ \mbox{and}  \  \  \  \
A^{-}\sim  \frac{\Delta^{N}d}{d}
\end{equation}
evaluated at $x=0.435$, at a c.m. energy  $\sqrt{s}=500
\mbox{GeV}$. So the region $y\sim -1$ is very sensitive to the
antiquark Sivers functions, whereas the region $y\sim +1$ is
sensitive to the quark Sivers functions. It is interesting to
notice the close correspondence between the expression for $A^{+}$
(see Eq. (8)) and the parity violating-asymmetry $A_L^{W^+}$
introduced in Ref. \cite{BS93}. Clearly, as well known,
$A_L^{W^{\pm}}$ allows the flavor separation of the quark helicity
distributions, similarly the measurement of $A_N^{W^{\pm}}$ is a
practical way to separate the $u$ and $d$ quarks Sivers functions
and their corresponding antiquarks $\bar u$ and $\bar d$. A
straightforward interpretation of a non-zero $A_L^{W^{\pm}}$ is,
in fact, a little bit more complicated because of its $p_T$ -
dependence, namely the factor $H(p_T)$ in Eq. (7), which is
unknown. It is possible to avoid this difficulty and to increase
statistics by integrating over the $p_T$ - range of the produced
$W^{\pm}$'s. This is part of the reason why we cannot make any
reliable prediction for these SSA, but the observation of
significant effects will be the unambiguous signature for the
presence of non-zero Sivers functions. On the experimental side,
we recall that a vast spin programme is under way at BNL-RHIC,
which will operate with polarized proton beams, both transversely
and longitudinally, up to $\sqrt{s}=500 \mbox{GeV}$ \cite{BSSV}.
All appropriate studies have been now completed in view of the
measurement of $A_L^{W^{\pm}}$, which will be precisely
determined, because the high luminosity of the machine allows a
copious production of $W^{\pm}$. So our proposal to measure
$A_N^{W^{\pm}}$ is pretty obvious and does not requiere any
further experimental effort. Another point worth mentioning is the
fact that if one considers prompt photon production
$pp^{\uparrow}\rightarrow \gamma X$, with a transversely polarized
proton, the SSA for this reaction, which does not involve any
fragmentation function, is also directly sensitive to Sivers
functions. Moreover, since the dominant subprocess is gluon
Compton scattering $g q \rightarrow \gamma q$, by selecting the
rapidity of the produced photon, it is possible to separate quark
the Sivers function from the gluon Sivers function, this last
object having not yet been considered in the literature. This
process has been also carefully studied for the determination of
the gluon helicity distribution at BNL-RHIC \cite{BSSV}.

To summarize, these SSA which are very well accessible at
BNL-RHIC, might open the way to a new class of relevant spin
effects at high enrgies, contrary to naive expectations.\\

{\bf Acknowledgments: } This work is partially supported by
Fondecyt (Chile) under Grant Number 1039355 and by the cooperation
programme Ecos-Conicyt C99E08 between France and Chile.

\end{document}